\documentclass[12pt]{article}

\usepackage{algorithm}				
\usepackage[noend]{algpseudocode}	
\usepackage{amsmath}				
\usepackage{amssymb}				
\usepackage{amsthm}				
\usepackage{caption}
\usepackage{subcaption}
\usepackage{color}
\usepackage[margin=1in]{geometry}
\usepackage[hidelinks]{hyperref}
\usepackage{titling}

\usepackage{complexity}
\usepackage{graphicx}
\usepackage{tikz}
\usetikzlibrary{matrix}

\theoremstyle{plain}				
\newtheorem{theorem}{Theorem}
\newtheorem{lemma}[theorem]{Lemma}

\theoremstyle{definition}			
\newtheorem{definition}[theorem]{Definition}

\theoremstyle{remark}			
\newtheorem*{remark}{Remark}

\newtheorem*{claim}{Claim}

\newcommand{\TDFA}{\textsf{2DFA-4W}}

\newcommand{\TDFATW}{\textsf{2DFA-3W}}

\newcommand{\TDFATWOW}{\textsf{2DFA-2W}}

\newcommand{\TNFA}{\textsf{2NFA-4W}}

\newcommand{\TNFATW}{\textsf{2NFA-3W}}

\newcommand{\TNFATWOW}{\textsf{2NFA-2W}}

\newcommand{\TDRATW}{\textsf{2DFA-3W}}
\newcommand{\TDRATWOW}{\textsf{2DFA-2W}}
\newcommand{\TNRATW}{\textsf{2NFA-3W}}
\newcommand{\TNRATWOW}{\textsf{2NFA-2W}}

\renewcommand{\phi}{\varphi}

\thanksmarkseries{alph}

\title{Degrees of Restriction for Two-Dimensional Automata}
\author{Taylor J. Smith \thanks{School of Computing, Queen's University, Kingston, Ontario, Canada. Email: \texttt{\{tsmith,ksalomaa\}@cs.queensu.ca}.} \and Kai Salomaa \thanksmark{1}}
\date{\today}

\begin{document}


\maketitle

\begin{abstract}
A three-way (resp., two-way) two-dimensional automaton has a read-only input head that moves in three (resp., two) directions on a finite array of cells labelled by symbols of the input alphabet. Restricting the input head movement of a two-dimensional automaton results in a model that is weaker in terms of recognition power.

In this paper, we introduce the notion of ``degrees of restriction" for two-dimensional automata, and we develop sets of extended two-dimensional automaton models that allow for some bounded number of restricted moves. We establish recognition hierarchies for both deterministic and nondeterministic extended three-way two-dimensional automata, and we find similar hierarchies for both deterministic and nondeterministic extended two-way two-dimensional automata. We also prove incomparability results between nondeterministic and deterministic extended three-way two-dimensional automata. Lastly, we consider closure properties for some operations on languages recognized by extended three-way two-dimensional automata.

\medskip

\noindent\textit{Key words and phrases:} closure properties, degrees of restriction, extended two-dimensional automata, recognition properties

\medskip

\noindent\textit{MSC2020 classes:} 68Q45 (primary); 68Q15 (secondary).
\end{abstract}


\section{Introduction}\label{sec:introduction}

The two-dimensional automaton model is a generalization of the one-dimensional (or string) automaton model that takes as input an array or matrix of symbols from some alphabet $\Sigma$. The input head of such an automaton can move either upward, downward, leftward, or rightward within its input word. The two-dimensional automaton model was introduced by Blum and Hewitt \cite{BlumHewitt19672DAutomata}.

If we restrict the input head movement of a two-dimensional automaton so that it cannot move in certain directions, then we obtain variants of the model that are weaker in terms of recognition power. If we restrict upward moves only, then we obtain what is known as a three-way two-dimensional automaton. Similarly, if we restrict both upward and leftward moves, then we obtain what is known as a two-way two-dimensional automaton. The three-way two-dimensional automaton model was introduced by Rosenfeld \cite{Rosenfeld1979PictureLanguages}, while the two-way two-dimensional automaton model was introduced by Anselmo et al.\ \cite{Anselmo2005NewOperations2DLanguages} and formalized by Dong and Jin \cite{Dong2012TwoWay2DAutomata}.

Motivated by the question of how the accepting power of two-dimensional computational models is affected by the number of input head reversals, Morita et al.~\cite{Morita2003InputHeadReversalBounded} studied ``input head reversal-bounded" two-dimensional Turing machines. In their paper, Morita et al.\ considered two-dimensional Turing machines operating on square tapes whose input heads may switch their vertical direction of movement some bounded number of times. The authors investigated a relationship between deterministic and nondeterministic input head reversal-bounded two-dimensional Turing machines, proposed a reversal hierarchy of space-bounded two-dimensional Turing machines, and established necessary and sufficient conditions for three-way two-dimensional Turing machines to simulate reversal-bounded four-way two-dimensional automata.

In the present paper, we consider a similar idea, which we term ``degrees of restriction" for two-dimensional automata. 
We introduce an $i$-extended three-way two-dimensional automaton model, $i \in \mathbb{N}$, where the computation on any input word may move downward, leftward, and rightward, and is additionally permitted to make at most $i$ upward moves. Similarly, we introduce an $(i,j)$-extended two-way two-dimensional automaton, $i,j \in \mathbb{N}$, which is permitted to make at most $i$ upward moves and at most $j$ leftward moves in a computation. 
The $i$ and $j$ bounds can be viewed as being maintained by a counter stored on an auxiliary tape that the automaton can only read from and decrement; this is to prevent the automaton from using the tape as general storage. Equivalently, the counter can be viewed as a stack containing some predefined number of unary symbols that can only be popped.

Automata with reversal-bounded counters have been studied in the past; for example, see Ibarra's survey~\cite{Ibarra2014AutomataReversalBounded}. However, the models defined in past work differ from our present model in that the counters of the former models may be either decremented or incremented. For simplicity, in our model, we assume that counters may only be decremented. Our model also differs from those considered by Morita et al.\ in that we do not restrict the number of input head reversals (e.g., down-to-up or up-to-down), but rather the number of restricted moves made by the input head (e.g., upward moves only). Thus, in our model, an automaton can make any number of normal input head moves, but only a limited number of restricted moves.


\section{Preliminaries}\label{sec:preliminaries}

A two-dimensional word consists of a finite array, or rectangle, of cells each labelled by a symbol from a finite alphabet $\Sigma$. When a two-dimensional word is written on the input tape of a two-dimensional automaton, the cells around the word are labelled by a special boundary marker $\# \not\in \Sigma$. A two-dimensional automaton has a finite state control that is capable of moving its input head in four directions within its input word: up, down, left, and right (denoted by $U$, $D$, $L$, and $R$, respectively).

\begin{definition}[Two-dimensional automaton]\label{def:2DFA}
A two-dimensional automaton is a tuple \allowbreak $(Q, \Sigma, \delta, q_{0}, q_{\rm accept})$, where $Q$ is a finite set of states, $\Sigma$ is the input alphabet (with $\# \not\in \Sigma$ acting as a boundary symbol), $\delta: (Q \setminus \{q_{\rm accept}\}) \times (\Sigma \cup \{\#\}) \to Q \times \{U, D, L, R\}$ is the partial transition function, and $q_{0}, q_{\rm accept} \in Q$ are the initial and accepting states, respectively.
\end{definition}

We can modify the deterministic model given in Definition~\ref{def:2DFA} to be nondeterministic by changing the transition function to map to $2^{Q \times \{U, D, L, R\}}$. We denote the deterministic and nondeterministic two-dimensional automaton models by \TDFA\ and \TNFA, respectively.

By restricting the movement of the input head, we obtain the aforementioned restricted variants of the two-dimensional automaton model. By prohibiting upward movements, we obtain the three-way two-dimensional automaton model. Similarly, by prohibiting both upward and leftward movements, we obtain the two-way two-dimensional automaton model.

\begin{definition}[Three-way/two-way two-dimensional automaton]
A three-way (resp., two-way) two-dimensional automaton is a tuple $(Q, \Sigma, \delta, q_{0}, q_{\rm accept})$ as in Definition~\ref{def:2DFA}, where the transition function $\delta$ is restricted to use only the directions $\{D, L, R\}$ (resp., the directions $\{D, R\}$).
\end{definition}

We denote deterministic and nondeterministic three-way two-dimensional automata by \TDFATW\ and \TNFATW, respectively, while the two-way model is denoted by the suffix \textsf{-2W}.

Additional details about the two-dimensional automaton model and its restrictions can be found in surveys by Inoue and Takanami \cite{Inoue19912DAutomataSurvey}, Kari and Salo \cite{KariSalo2011PictureWalkingAutomataSurvey}, and the first author \cite{Smith2019TwoDimensionalAutomata}.

We now move on to defining the main models of the paper, which we call ``extended" two-dimensional automata. We denote a (deterministic) $i$-extended three-way two-dimensional automaton by $\TDRATW[i]$, where $i$ is the number of upward moves the input head is permitted to make. Similarly, we denote a (deterministic) $(i,j)$-extended two-way two-dimensional automaton by $\TDRATWOW[i, j]$, where $i$ (resp., $j$) is the number of upward (resp., leftward) moves the input head is permitted to make.

Clearly, $\TDRATWOW[\infty, \infty] = \TDRATW[\infty] = \TDFA$. Similarly, $\TDRATWOW[0, \infty] = \TDFATW$, and $\TDRATWOW[\infty, 0]$ is equivalent to $\TDFATW^{\circlearrowleft}$, 
or the class of deterministic three-way two-dimensional automata where the transition function is restricted to use only the directions $\{U, D, R\}$. 
Lastly, $\TDRATW[0] = \TDFATW$ and $\TDRATWOW[0, 0] = \TDFATWOW$. All of the previous definitions and results apply also to nondeterministic models.


\section{Recognition Properties}\label{sec:recognition}


\subsection{Three-Way Recognition}

We begin by examining relationships between three-way and four-way two-dimensional automata and the $i$-extended three-way variant.

\begin{theorem}\label{thm:TNRATWzerosubsetone}
$\TNRATW[0] \subset \TNRATW[1]$.

\begin{proof}
Let $\Sigma = \{\texttt{0}, \texttt{1}\}$. We define the language $L_{1}$ to be the language of all two-dimensional words $w$ with two rows such that $w$ contains at least two occurrences of ``stacked \texttt{1}s"; that is, at least two columns $j$ where $w[1,j] = w[2,j] = \texttt{1}$. An example of a word in the language $L_{1}$ is illustrated in Figure~\ref{fig:stackedword}.

\begin{figure}[t]
\[\arraycolsep=2pt\def\arraystretch{0.95}
\begin{array}{ccccccccccccc}
\#		& \#		& \#		& \#		& \#		& \#		& \#		& \#		& \#		& \#		& \#		& \#		& \# \\
\#		& \ast	& \cdots	& \ast	& \texttt{1}	& \ast	& \cdots	& \ast	& \texttt{1}	& \ast	& \cdots	& \ast	& \# \\
\#		& \ast	& \cdots	& \ast	& \texttt{1}	& \ast	& \cdots	& \ast	& \texttt{1}	& \ast	& \cdots	& \ast	& \# \\
\#		& \#		& \#		& \#		& \#		& \#		& \#		& \#		& \#		& \#		& \#		& \#		& \# \\
\end{array}
\]
\caption{An example of a word in the language $L_{1}$ from the proof of Theorem~\ref{thm:TNRATWzerosubsetone}, where the symbol $\ast$ denotes either \texttt{0} or \texttt{1}}
\label{fig:stackedword}
\end{figure}

An automaton $\mathcal{A} \in \TNRATW[1]$ recognizes words in $L_{1}$ via the following procedure:
\begin{enumerate}
\item The input head of $\mathcal{A}$ moves rightward and scans the first row of the input word until it nondeterministically selects an occurrence of \texttt{1}.
\item The input head moves downward to verify that the symbol in the second row is a \texttt{1}.
\item The input head moves rightward and scans the second row of the input word until it nondeterministically selects another occurrence of \texttt{1}.
\item\label{itm:L1step4} The input head moves upward to verify that the symbol in the first row is a \texttt{1}.
\end{enumerate}
Clearly, this procedure requires only one upward move, which is performed in Step~\ref{itm:L1step4}.

Recall that any automaton $\mathcal{A}' \in \TNRATW[0]$ is in fact a three-way two-dimensional automaton. For the sake of contradiction, suppose such an automaton $\mathcal{A}'$, with $m$ states, recognizes the language $L_{1}$.

Let $u(i, j, z)$ denote the single-row word (i.e., one-dimensional word) of length $z$ where cells at positions $i$ and $j$ contain the symbol \texttt{1}, $1 \leq i < j \leq z$, and all other positions contain the symbol \texttt{0}. Furthermore, let $w(i, j, z)$ denote the two-row word where both rows are exactly the word $u(i, j, z)$. This two-dimensional word contains exactly two occurrences of ``stacked \texttt{1}s", while all other cells contain \texttt{0}s.

Choose a length $z$ such that $(z-1) / 2 > m$. The automaton $\mathcal{A}'$ must accept all words $w(i, j, z)$, where $1 \leq i < j \leq z$. For each such word, let $C(i, j, z)$ denote an accepting computation of $\mathcal{A}'$ on that word. Since $z(z - 1) / 2 > m \cdot z$, there exist two different computations $C(i, j, z)$ and $C(r, s, z)$, with $i \neq r$ or $j \neq s$, such that in both computations $C(i, j, z)$ and $C(r, s, z)$ the automaton $\mathcal{A}'$ makes a downward move to the second row in the same column and same state. Without loss of generality, all accepting computations of $\mathcal{A}'$ must make a downward move, since otherwise all symbols on the second row of the input word could be \texttt{0}.

However, as a consequence of this observation, $\mathcal{A}'$ will also accept the two-dimensional word $w_{0}$ where the first row is the word $u(i, j, z)$ and the second row is $u(r, s, z)$. Since $i \neq r$ or $j \neq s$, the word $w_{0}$ contains at most one occurrence of ``stacked \texttt{1}s" and, therefore, is not in the language $L_{1}$.
\end{proof}
\end{theorem}

We can use a similar argument to generalize the previous theorem to work for any number of upward moves $i$.

\begin{theorem}\label{thm:TNRATWgeneral}
$\TNRATW[i] \subset \TNRATW[i + 1]$.

\begin{proof}
Recall the language $L_{1}$ from the proof of Theorem~\ref{thm:TNRATWzerosubsetone}. We create a family of languages $L_{i}$, $i \geq 2$, by taking $i$ copies of $L_{1}$ and concatenating row-wise to form a new language. In this way, we create a language of two-dimensional words each consisting of $2i$ rows where rows $2j$ and $2j + 1$, $0 \leq j < i$, contain at least two occurrences of ``stacked \texttt{1}s" as defined earlier.

An automaton $\mathcal{B} \in \TNRATW[i+1]$ recognizes words in $L_{i+1}$ via the following procedure:
\begin{enumerate}
\item The input head of $\mathcal{B}$ follows the process presented in the proof of Theorem~\ref{thm:TNRATWzerosubsetone} to verify that the first two rows of the input word contain at least two occurrences of ``stacked \texttt{1}s".
\item After verifying that the first two rows satisfy this condition, the input head moves back to the left border of the input word and then makes two downward moves.
\item The automaton repeats these two steps until all consecutive pairs of rows are checked.
\end{enumerate}
The first two steps of this procedure require one upward move, and since there are a total of $2i + 2$ rows in the input word, the procedure must be repeated $i + 1$ times. Therefore, in order for $\mathcal{B}$ to accept an input word, it must make a total of $i + 1$ upward moves.

For the sake of contradiction, suppose an automaton $\mathcal{B}' \in \TNRATW[i]$ with $m$ states also recognizes words in $L_{i+1}$. 
Each word in $L_{i+1}$ consists of $2i + 2$ rows where, for each odd $c$, the $c$th and $(c+1)$st rows form a word in $L_{1}$. Choose the number of columns $z$ such that
\begin{equation}\label{eq:zed}
(z - 1)/2 > m \cdot (i + 1).
\end{equation}
Let $v(j, k, z)$ denote a two-dimensional word with $z$ columns each of length $2i + 2$, where the $j$th and $k$th columns, $1 \leq j < k \leq z$, consist entirely of \texttt{1}s and all other symbols are \texttt{0}. Let $C(j, k, z)$ denote an accepting computation of $\mathcal{B}'$ on the word $v(j, k, z)$.

Since $C(j, k, z)$ can make at most $i$ upward moves, by the choice of $z$ in Equation~\ref{eq:zed}, there exist $C(j, k, z)$ and $C(r, s, z)$, with $(j,k) \neq (r,s)$, such that for some even row value $x$, both $C(j, k, z)$ and $C(r, s, z)$ do not make an upward move from row $x$ to row $x-1$ and both $C(j, k, z)$ and $C(r, s, z)$ make a downward move from row $x-1$ to row $x$ in the same column and same state.

However, as a consequence of this observation, $\mathcal{B}'$ must accept a word where the first $x-1$ rows are $v(j, k, z)$ and the last $2i + 3 - x$ rows are $v(r, s, z)$. Since $j \neq r$ and $k \neq s$, the $(x-1)$st and $x$th rows do not form a word in $L_{1}$.
\end{proof}
\end{theorem}

Combining Theorems~\ref{thm:TNRATWzerosubsetone} and \ref{thm:TNRATWgeneral} together with the fact that, for all $i \geq 1$, $\TNRATW[i] \subset \TNRATW[\infty] = \TNFA$, we obtain 
a recognition hierarchy amongst nondeterministic extended three-way two-dimensional automata. 
A similar hierarchy exists for deterministic models.

\begin{theorem}\label{thm:TDRATWhierarchy}
For all $i \geq 1$,
\begin{equation*}
\TDFATW \subset \cdots \subset \TDRATW[i] \subset \TDRATW[i + 1] \subset \cdots \subset \TDFA.
\end{equation*}
\end{theorem}

The proof of Theorem~\ref{thm:TDRATWhierarchy} goes through in a similar way to other proofs in this section, but it uses a different language $M_{i}$ to separate different classes in the hierarchy. 
We define the language $M_{1}$ to be the language of all two-dimensional words with two rows where each row contains exactly two occurrences of ``stacked \texttt{1}s" as defined earlier\footnote{Note that the language $M_{1}$ differs from the language $L_{1}$ in that words in $M_{1}$ contain \emph{exactly} two occurrences of ``stacked \texttt{1}s", while words in $L_{1}$ contain \emph{at least} two occurrences of ``stacked \texttt{1}s".}. We can then create a family of languages $M_{i}$, $i \geq 2$, by taking $i$ copies of $M_{1}$ and concatenating row-wise.


\subsection{Deterministic vs.\ Nondeterministic Three-Way Recognition}

Thus far, we have established that separate hierarchies exist for deterministic and nondeterministic $i$-extended three-way two-dimensional automata. When we compare deterministic and nondeterministic models to each other, however, it turns out that the models are incomparable. This stands in contrast to the usual relationship between deterministic and nondeterministic three-way two-dimensional automata, where $\TDFATW \subset \TNFATW$ \cite{Rosenfeld1979PictureLanguages}. 
In what follows, we present a series of lemmas building up to the main result of this section.

\begin{lemma}\label{lem:TDFATW1vsTNFATW0}
There exists a language $M_{1}$ that is recognized by an automaton in $\TDFATW[1]$ and not recognized by any automaton in $\TNFATW[0]$.

\begin{proof}
Let $\Sigma = \{\texttt{0}, \texttt{1}\}$. Recall the definition of the language $M_{1}$ from the discussion following Theorem~\ref{thm:TDRATWhierarchy}. We shall use this language again in the present proof.

An automaton $\mathcal{M} \in \TDFATW[1]$ recognizes $M_{1}$ via the following procedure:
\begin{enumerate}
\item The input head of $\mathcal{M}$ scans the first row of its input word and verifies that the row contains exactly two occurrences of \texttt{1}.
\item The input head returns to the leftmost occurrence of \texttt{1} in the first row and makes one downward move to verify that the symbol in the corresponding column of the second row is also \texttt{1}.
\item The input head scans the second row of its input word and verifies that the row contains exactly two occurrences of \texttt{1}.
\item The input head returns to the rightmost occurrence of \texttt{1} in the second row and makes one upward move to verify that the symbol in the corresponding column of the first row is also \texttt{1}.
\end{enumerate}
Clearly, this procedure requires only one upward move.

However, an automaton $\mathcal{N} \in \TNFATW[0]$ cannot recognize any words in the language $M_{1}$ by an argument analogous to that given in the proof of Theorem~\ref{thm:TNRATWzerosubsetone} showing that no automaton $\mathcal{B} \in \TNFATW[0]$ can recognize words in the language $L_{1}$.
\end{proof}
\end{lemma}

We may generalize the previous argument to apply to $i$-extended three-way two-dimensional automata for any value of $i \geq 1$.

\begin{lemma}\label{lem:TDFATWivsTNFATWi}
There exists a language $M_{i+1}$ that is recognized by an automaton in $\TDFATW[i+1]$ and not recognized by any automaton in $\TNFATW[i]$.

\begin{proof}
Recall the family of languages $M_{i}$, $i \geq 2$, from the discussion following Theorem~\ref{thm:TDRATWhierarchy}. We shall use these languages again in the present proof.

An automaton $\mathcal{M}' \in \TDFATW[i+1]$ recognizes words in $M_{i+1}$ via the following procedure:
\begin{enumerate}
\item The input head of $\mathcal{M}'$ follows the process presented in the proof of Lemma~\ref{lem:TDFATW1vsTNFATW0} to verify that the first two rows of the input word contain exactly two occurrences of ``stacked \texttt{1}s".
\item After verifying that the first two rows satisfy this condition, the input head moves back to the left border of the input word and then makes two downward moves.
\item The automaton repeats these two steps until all consecutive pairs of rows are checked.
\end{enumerate}
Since this procedure must be repeated $i+1$ times, $\mathcal{M}'$ makes a total of $i+1$ upward moves.

However, an automaton $\mathcal{N}' \in \TNFATW[i]$ cannot recognize any words in the language $M_{i+1}$ by an argument analogous to that given in the proof of Theorem~\ref{thm:TNRATWgeneral} showing that no automaton $\mathcal{B}' \in \TNFATW[i]$ can recognize words in the language $L_{i+1}$.
\end{proof}
\end{lemma}

Next, we consider the opposite direction.

\begin{lemma}\label{lem:TNFATW1vsTDFATW0}
There exists a language $N_{2}$ that is recognized by an automaton in $\TNFATW[0]$ and not recognized by any automaton in $\TDFATW[1]$.

\begin{proof}
Let $\Sigma = \{\texttt{0}, \texttt{1}\}$. We define the language $N_{1}$ to be the language of all two-dimensional words consisting of two rows that contain at least one occurrence of ``stacked \texttt{1}s". We then define $N_{2}$ to be the language created by concatenating two copies of $N_{1}$ row-wise; that is, $N_{2}$ is the language of all two-dimensional words consisting of four rows, where the first two rows and the last two rows each contain at least one occurrence of ``stacked \texttt{1}s".

An automaton $\mathcal{P} \in \TNFATW[0]$ recognizes words in $N_{2}$ via the following procedure:
\begin{enumerate}
\item The input head of $\mathcal{P}$ moves rightward and scans the first row of the input word until it nondeterministically selects an occurrence of \texttt{1}.
\item The input head moves downward to verify that the symbol in the second row is a \texttt{1}.
\item The input head moves to the leftmost symbol, makes a downward move to the third row of the input word, and scans the third row until it nondeterministically selects another occurrence of \texttt{1}.
\item The input head moves downward to verify that the symbol in the fourth row is a \texttt{1}.
\end{enumerate}
Moreover, this procedure does not require any upward moves.

To show that no automaton $\mathcal{Q} \in \TDFATW[1]$ is capable of recognizing words in $N_{2}$, we first prove an intermediate claim.

\begin{claim}
No automaton $\mathcal{Q}' \in \TDFATW[0]$ is capable of recognizing words in $N_{1}$.

\begin{proof}
For the sake of contradiction, suppose that an automaton $\mathcal{Q}' \in \TDFATW[0]$ with $t$ states recognizes words in $N_{1}$. Select a value $z$ such that $(z - 1)/2 > t$. Let $x(i, j, z)$ denote the single-row word (i.e., one-dimensional word) of length $z$ where cells at positions $i$ and $j$ contain the symbol \texttt{1}, $0 \leq i < j \leq z$, and all other positions contain the symbol \texttt{0}.

When $\mathcal{Q}'$ reads a word $x(i, j, z)$, it must eventually move downward, since it must accept some words in which $x(i, j, z)$ is the first row. Since $z(z-1)/2 > t \cdot z$, there exist pairs $(i, j)$ and $(r, s)$, with $(i, j) \neq (r, s)$, such that when operating on the rows $x(i, j, z)$ and $x(r, s, z)$, the input head of $\mathcal{Q}'$ moves downward in the same position and same state.

\begin{remark}
Note that the pair $(i, j)$ contains at least one element that does not appear in the pair $(r, s)$. This is because we have chosen the two pairs in such a way that $i \neq r$ or $j \neq s$. In the case where $i \neq r$, if $i = s$, then $j > i$ does not appear in the pair $(r, s)$. The other case is symmetric.
\end{remark}

We assume that $i$ does not appear in the pair $(r, s)$; that is, $i \neq r$ and $i \neq s$. By our assumption that $\mathcal{Q}'$ recognizes words in $N_{1}$, $\mathcal{Q}'$ must accept the two-dimensional word whose first row is $x(i, j, z)$ and whose second row is of the form $\texttt{0}^{i-1}\texttt{1}\texttt{0}^{z-i}$. Since $\mathcal{Q}'$ is deterministic, and from our earlier observation that the input head of $\mathcal{Q}'$ moves downward in the same position and same state on some rows $x(i, j, z)$ and $x(r, s, z)$, it follows that $\mathcal{Q}'$ must also accept the two-dimensional word whose first row is $x(r, s, z)$ and whose second row is of the form $\texttt{0}^{i-1}\texttt{1}\texttt{0}^{z-i}$. However, since $i \neq r$ and $i \neq s$, this word is not in $N_{1}$.
\end{proof}
\end{claim}

Suppose that an automaton $\mathcal{Q} \in \TDFATW[1]$ recognizes words in $N_{2}$. Using an argument similar to that used in the preceding claim, $\mathcal{Q}$ cannot correctly verify that the first two rows of its input word form a word from $N_{1}$ without making an upward move. If, for all input words, $\mathcal{Q}$ reads the first two rows without making an upward move, then it will enter the third row of an input word whose first two rows do not form a word from $N_{1}$. If we suppose the third and fourth rows of this input word consist entirely of \texttt{1}s, then this forces $\mathcal{Q}$ to accept an illegal input word.

Thus, for some input words, $\mathcal{Q}$ must make an upward move in the first two rows. Again, using an argument similar to that used in the preceding claim, it follows that $\mathcal{Q}$ cannot correctly verify that the third and fourth rows of its input word form a word from $N_{1}$. Therefore, if $\mathcal{Q}$ accepts all words in $N_{2}$, it must also accept some words not in $N_{2}$.
\end{proof}
\end{lemma}

As a consequence of Lemmas~\ref{lem:TDFATW1vsTNFATW0} and \ref{lem:TNFATW1vsTDFATW0}, we obtain the aforementioned result.

\begin{theorem}\label{thm:TDFATW1vsTNFATW0incomparable}
The classes $\TNFATW[0]$ and $\TDFATW[1]$ are incomparable.
\end{theorem}


\subsection{Two-Way Recognition}

In this section, we turn to examining relationships between two-way and three-way two-dimensional automata and the $(i,j)$-extended two-way variant.

We prove our first result relating to extended two-way two-dimensional automata in a manner similar to the three-way case presented in Theorem~\ref{thm:TNRATWzerosubsetone}.

\begin{theorem}\label{thm:TNRATWOWzerosubsetone}
$\TNRATWOW[0,0] \subset \TNRATWOW[1,0]$.

\begin{proof}
Let $\Sigma = \{\texttt{0}, \texttt{1}\}$. Recall the language $L_{1}$ from the proof of Theorem~\ref{thm:TNRATWzerosubsetone}.

An automaton $\mathcal{C} \in \TNRATWOW[1,0]$ recognizes words in $L_{1}$ in exactly the same way as the automaton $\mathcal{A} \in \TNRATW[1]$ from the proof of Theorem~\ref{thm:TNRATWzerosubsetone}. Again, this process requires only one upward move, and no leftward moves are needed.

Recall that any automaton $\mathcal{C}' \in \TNRATWOW[0,0]$ is in fact a two-way two-dimensional automaton. For the sake of contradiction, suppose such an automaton $\mathcal{C}'$ recognizes the language $L_{1}$. Then $\mathcal{C}'$ must necessarily accept the word
\begin{equation*}
\begin{array}{cc}
\texttt{1} & \texttt{1} \\
\texttt{1} & \texttt{1}
\end{array}.
\end{equation*}
Moreover, the accepting computation of $\mathcal{C}'$ on this word cannot visit all cells of the word. Thus, if we change the one unvisited cell to contain a \texttt{0} instead of a \texttt{1}, then $\mathcal{C}'$ must also accept this word not in $L_{1}$.
\end{proof}
\end{theorem}

We now move on to proving the generalized form of the previous theorem. The idea of the proof is similar to that used in Theorem~\ref{thm:TNRATWgeneral}, but due to the fact that the model under consideration cannot move leftward, we must consider a simplified version of the language $L_{i + 1}$ that does not require an automaton to make leftward moves in order to accept words from the language.

\begin{theorem}\label{thm:TNRATWOWgeneral}
$\TNFATWOW[i,0] \subset \TNFATWOW[i+1, 0]$.

\begin{proof}
Let $\Sigma = \{\texttt{0}, \texttt{1}\}$. We define the language $K_{i+1}$ to be the language of all two-dimensional words $u$ with two rows such that $u$ contains at least $2i + 2$ occurrences of ``stacked \texttt{1}s".

An automaton $\mathcal{D} \in \TNRATWOW[i+1,0]$ recognizes words in $K_{i+1}$ via the following procedure:
\begin{enumerate}
\item The input head of $\mathcal{D}$ moves rightward and scans the first row of the input word until it nondeterministically selects the first occurrence of ``stacked \texttt{1}s". The input head then moves downward to verify the other occurrence of \texttt{1}.
\item The input head moves rightward and scans the second row of the input word until it nondeterministically selects the second occurrence of ``stacked \texttt{1}s". The input head then moves upward to verify the other occurrence of \texttt{1}.
\item The input head repeats the previous two steps a total of $i+1$ times to verify that a total of $2i+2$ columns of the input word contain ``stacked \texttt{1}s".
\end{enumerate}
Altogether, this procedure requires a total of $i+1$ upward moves, and no leftward moves.

For the sake of contradiction, suppose an automaton $\mathcal{D}' \in \TNRATWOW[i,0]$ recognizes the language $K_{i+1}$. Then $\mathcal{D}'$ must accept the two-dimensional word $u_{0}$ consisting of two rows, where each row consists of the string $\texttt{1}^{2i+2}$. However, $\mathcal{D}'$ cannot visit all cells of $u_{0}$, since it can only make $i$ upward moves. Moreover, $\mathcal{D}'$ cannot backtrack through the word, as it cannot make any leftward moves. Thus, if we change one unvisited cell in $u_{0}$ to contain a \texttt{0} instead of a \texttt{1}, then $\mathcal{D}'$ must also accept this word not in $K_{i+1}$.
\end{proof}
\end{theorem}

Combining Theorems~\ref{thm:TNRATWOWzerosubsetone} and \ref{thm:TNRATWOWgeneral} together with the fact that, for all $i \geq 1$, $\TNRATWOW[i,0] \subset \TNRATWOW[\infty, 0] = \TNFATW^{\circlearrowleft}$, we obtain another recognition hierarchy for the two-way case. 
Naturally, we get a similar hierarchy as before for deterministic models.

\begin{theorem}\label{thm:TDFATWOWhierarchy}
For all $i \geq 1$,
\begin{equation*}
\TDFATWOW \subset \cdots \subset \TDRATWOW[i, 0] \subset \TDRATWOW[i + 1, 0] \subset \cdots \subset \TDFATW^{\circlearrowleft}.
\end{equation*}
\end{theorem}

To prove Theorem~\ref{thm:TDFATWOWhierarchy}, we use the singleton language $S_{i}$ over the alphabet $\Sigma = \{\texttt{0}, \texttt{1}\}$, which consists of one word of dimension $2 \times i$ where all symbols are \texttt{1}. Then, in a manner similar to the proof of Theorem~\ref{thm:TNRATWOWgeneral}, we can show that a deterministic two-way two-dimensional automaton making $(i+1)$ upward moves can recognize the word in $S_{2i+2}$, while no deterministic two-way two-dimensional automaton making $i$ upward moves can visit all cells in such a word.

Since the classes $\TDFATW/\TNFATW$ and $\TDFATW^{\circlearrowleft}/\TNFATW^{\circlearrowleft}$ are equivalent up to rotation\footnote{By ``up to rotation", we mean that a language $L$ recognized by an automaton $\mathcal{A} \in \TNFATW^{\circlearrowleft}$ can also be recognized by an automaton $\mathcal{A}' \in \TNFATW$ if each word in $L$ is rotated clockwise by 90 degrees. The same applies to deterministic models.}, both of the two-way hierarchies 
are upper-bounded by the class of languages recognized by traditional three-way two-dimensional automata.

Moreover, if a two-way two-dimensional automaton can recognize a language using at most $i$ upward moves and no leftward moves, then clearly the ``transpose" of that language (i.e., each word in that language reflected about its diagonal) can be recognized by a two-way two-dimensional automaton using no upward moves and at most $i$ leftward moves. Therefore, there exist analogous hierarchies  
for the $(0,i)$-extended models $\TNRATWOW[0,i]$ and $\TDRATWOW[0,i]$, where $i \geq 1$.


\section{Closure Properties}\label{sec:closure}

In this section, we take a brief diversion to investigate some closure properties for extended two-dimensional automata. 
From past work~\cite{InoueTakanami1979TapeBounded2DTuring, Sipser1980HaltingSpaceBounded}, we know that the classes of languages recognized by nondeterministic three-way and nondeterministic four-way two-dimensional automata are closed under the operations of union and reversal (or ``row reflection"). On the other hand, the deterministic equivalents of these language classes are closed only under language complement~\cite{Szepietowski1992SomeRemarks2DAutomata}. 
Thus, it is worthwhile to determine whether closure holds for the ``in-between" models of $\TDRATW[i]$ and $\TNRATW[i]$.

It seems clear that 
the class $\TNRATW[i]$ is closed under union, since we can simply take the set of automata recognizing each language in the union and nondeterministically choose which automaton to use on a given input word.

\begin{theorem}
For all $i \geq 1$, the class $\TNRATW[i]$ is closed under union.
\end{theorem}

Using an approach based on that used to prove closure for the traditional three-way two-dimensional automaton model, we can show that deterministic $i$-extended three-way two-dimensional automata are closed under complement.

\begin{theorem}
For all $i \geq 1$, the class $\TDRATW[i]$ is closed under complement.

\begin{proof}
Let $\mathcal{A} \in \TDRATW[i]$. We show that there exists an automaton $\mathcal{A}' \in \TDRATW[i]$ such that, if $\mathcal{A}$ accepts some input word, $\mathcal{A}'$ does not, and vice versa.

If $\mathcal{A}$ halts on every input word, then we simply swap the accepting and non-accepting states of $\mathcal{A}$ to obtain the automaton $\mathcal{A}'$. Otherwise, $\mathcal{A}$ loops infinitely on some input word, and we must show that $\mathcal{A}'$ can simulate the computation of $\mathcal{A}$ while ensuring that, in each row of its input word, the input head of $\mathcal{A}'$ either moves upward or downward, or the computation halts in that row.

To do so, $\mathcal{A}'$ simulates the computation of $\mathcal{A}$ in the style of Theorem 1 of Szepietowski~\cite{Szepietowski1992SomeRemarks2DAutomata}. The details of the construction are largely similar, with the exception of how right crossing sequences are computed. In our simulation, right crossing sequences may now account for a new outcome, $\{\uparrow\}$, corresponding to the case where the input head of $\mathcal{A}$ leaves the current row by making an upward move. This outcome is in addition to the existing outcomes given by Szepietowski, $\{\ell, \downarrow, \leftarrow\}$, which correspond to looping within the current row, leaving the current row by moving downward, and moving leftward beyond the leftmost boundary of the crossing sequence, respectively. We make an analogous change to the method of computing left crossing sequences. Since $\mathcal{A}$ can only make a limited number of upward moves, $\mathcal{A}'$ will never enter a ``vertical loop" (i.e., a loop where the input head of $\mathcal{A}$ returns infinitely often to an earlier row via upward moves). Thus, we need only to detect and handle loops within a single row, which is done by our modified form of Szepietowski's construction. Lastly, $\mathcal{A}'$ accepts if and only if $\mathcal{A}$ does not accept, and vice versa.
\end{proof}
\end{theorem}

Going further, an interesting set of language operations to study would be those operations that are closed for either three- or four-way two-dimensional automata, but not both. For instance, union, intersection, reversal, and rotation are closed for \TDFA, but not for \TDFATW. Similarly, intersection and rotation are closed for \TNFA\ but not for \TNFATW, while row concatenation and row closure are closed for \TNFATW\ but not for \TNFA. (Full details may be found in the surveys on two-dimensional automata~\cite{Inoue19912DAutomataSurvey, KariSalo2011PictureWalkingAutomataSurvey, Smith2019TwoDimensionalAutomata}.) Using our model, we could determine whether an operation becomes closed at some intermediate stage, or whether any modification to input head movement results in loss of closure.


\section{Conclusion}

Restricting the input head movement of a two-dimensional automaton results in a model that is weaker in terms of recognition power. However, based on past related work, it is reasonable to assume that this recognition power is affected by permitting some bounded number of input head reversals. In this paper, we considered the notion of degrees of restriction, and we developed extended two-dimensional automaton models in both three-way and two-way variants. We established that separate strict recognition hierarchies exist for both deterministic and nondeterministic $i$-extended three-way two-dimensional automata, and similar hierarchies exist for the two-way model. When we consider deterministic and nondeterministic extended three-way two-dimensional automata together, however, we find that the two models are incomparable. This is in contrast to the usual strict containment relationship between deterministic and nondeterministic three-way two-dimensional automata. We also investigated closure properties of extended three-way two-dimensional automata, finding that the nondeterministic model is closed under union and the deterministic model is closed under complement.

There remain some natural avenues for further study on this model. For the extended two-way two-dimensional automaton model, it would be worthwhile to investigate what kind of ``sub-hierarchy" might result when modifying the numbers of permitted upward and leftward moves simultaneously, rather than separately. There also remains the problem of establishing a relationship between $\TDFATWOW[i,j]$ and $\TNFATWOW[i,j]$, similar to the result of Theorem~\ref{thm:TDFATW1vsTNFATW0incomparable}. Moreover, the question of closure status persists for some operations such as reversal for $\TNFATW[i]$.

Lastly, in the introduction, we mentioned related work of Morita et al.\ on input-head reversal-bounded two-dimensional Turing machines, and we contrasted our model with theirs. We may alternatively consider a three-way two-dimensional automaton model more closely related to the model of Morita et al., where the automaton has two modes of operation 
(specifically, where its vertical direction of movement is either downward or upward) 
and it may switch between these two modes of operation some constant $k$ number of times. Investigating such a model may prove to be interesting in its own right, and could lead to results similar to those presented in this paper.


\bibliographystyle{plain}
\bibliography{References}


\end{document}